\documentstyle[aps]{revtex}
\begin{document}
\draft
\title{
PION STRUCTURE FUNCTION $F_{2}^{\pi}$ In the Valon Model
}

\author{Firooz Arash$^{(a,b)}$\footnote{e-mail: farash@cic.aut.ac.ir}}
\address{
$^{(a)}$ Center for Theoretical Physics and
Mathematics, AEOI P.O.Box 11365-8486, Tehran, Iran \\
$^{(b)}$
Physics Department, AmirKabir University (Tafresh
Campus), Hafez Avenue, Tehran, Iran \\
}
\date{\today}
\maketitle
\begin{abstract}
Partonic structure of constituent quark (or{\it{valon}}) in the
Next-to-Leading Order is used to calculate pion structure function.
This is a further demonstration of the finding that the constituent
quark structure is universal, and once it is calculated, the
Structure of any hadron can be predicted thereafter, using a convolution
method, without introducing any new free parameter. The results are compared with
the pion structure function from ZEUS Coll. Leading Neutron Production
in $e^{+}p$ collisions at HERA. We found good agreement with the experiment.
A resolution for the issue of normalization of the experimental data is 
suggested. In addition, the proportionality of $F_{2}^{\pi}$ and $F_{2}^{p}$,
which have caused confusion in the normalization of ZEUS data is discussed
and resolved.\\
{\bf PCAC}
\end{abstract}
\section{INTRODUCTION}
ZEUS Collaboration at HERA has recently published \cite{1} data on pion
structure function, $F_{2}^{\pi}$, using the leading neutron production in
$e^{+}p$ collision. The data do suggest that there is a simple relation
between proton structure function, $F_{2}^{p}$, and the pion structure function
$F_{2}^{\pi}$. This assertion is believable and points to the direction that
there exists a more basic and universal structure inside all hadrons.\\
In Ref.[1] the normalization of $\sigma_{\gamma\pi}$ and hence,
the pion structure function is fixed by two different methods: (a)
dominance of one pion exchange, and (b) use of the additive quark
model. The two normalizations differ by a factor of two. The
additive quark model makes no statement about the leading baryon
production while the extraction of $F_{2}^{\pi}$ is completely
based on meson exchange dynamics. A criticism of this procedure is
given in \cite{2}. It is evident that the issue of normalization
of the data is more uncertain than it was thought before. ZEUS Coll. now
believes that most likely the final result will lie between the
two options \cite{3}. The issue is what to use for $F(t)$ that
parameterizes the shape of the pion cloud in the proton. In this
paper we will offer an alternative, which does not rely on those
assumptions and renders support to a normalization, which lies
between the two options used by the ZEUS Coll. \\
In what follows we have used the structure function of a Constituent Quark (CQ),
(in Ref.\cite{4} it is called {\it{valon}} and hence the {\it{valon model}})
which is universal to all hadrons and from there, with a convolution method,
the structure function of pion is obtained. The motivation for such an
approach is based on the fact that such a model for soft production has found
phenomenological success at $\sqrt{s}<100$ GeV and low $p_{T}$. It should, however,
be understood that when $s$ is high enough to generate a significant component
of hard subprocesses $(\sqrt{s}>200 GeV)$, $\sigma_{tot}$ and average $p_{T}$ will
both increase and inclusive distributions will lose their scaling behavior.
Nevertheless, the soft component is unchanged, and the model remains valid.\\
Our knowledge of hadronic structure is based on the hadron
spectroscopy and the Deep Inelastic Scattering (DIS) data. In the
former picture quarks are massive particles and their bound states
describe the static properties of hadrons; while the
interpretation of DIS data relies upon the QCD Lagrangian, where
the hadronic structure is intimately connected with the presence
of a large number of partons. It has been shown \cite{5} that it
is possible to perturbatively dress a valence QCD Lagrangian field
to all order and construct a constituent quark in conformity with
the color confinement. Therefore, the assumption of constituent
quark as a valence quark with its cloud of partons is reasonable.
The Cloud is generated by QCD dynamics. A complete description and
calculational procedure for obtaining the constituent quark
structure, and hence, the hadronic structure functions are
detailed in \cite{6}.
\section{FORMALISM}
By definition, a valon is a universal building block for every hadron.
In a DIS experiment at high enough $Q^{2}$ it is the structure of
valon that is being probed, while at low enough $Q^{2}$ this structure cannot be
resolved and it behaves as the valence quark and the hadron is viewed as
the bound state of its valons. For a U-type valon one can write its structure
as,
\begin{equation}
F_{2}^{U}(z,Q^2)=\frac{4}{9}z(q_{\frac{u}{U}}+q_{\frac{\bar{u}}{U}})+ \frac{1}
{9}z(q_{\frac{d}{U}}+q_{\frac{\bar{d}}{U}}+q_{\frac{s}{U}}+q_{\frac{\bar{s}}{U}})+...
\end{equation}
where all the functions on the right-hand side are the probability functions
for quarks having momentum fraction $z$ of a U-type valon at $Q^{2}$.
These functions are calculated in Ref. [6] in the next-to-leading order and we
will not go into the details here. Suffices to note that the functional forms
of the parton distributions in a constituent quark, (or valon), is as follows [6]:
\begin{equation}
zq_{\frac{val.}{CQ}}(z,Q^{2})=a z^{b}(1-z)^{c}
\end{equation}
\begin{equation}
zq_{\frac{sea}{CQ}}(z,Q^{2}) = \alpha z^{\beta}(1-z)^{\gamma}[1+\eta z +\xi z^{0.5}]
\end{equation}
The parameters $a$, $b$, $c$, $\alpha$, etc. are functions of
$Q^{2}$ and are given in the appendix of Ref.[6]. The above
parameterization of the parton distribution in a valon is for light
quarks, $u$ and $d$. For heavy quarks additional phenomenological
assumptions are needed to be made. It is known that in proton the
strange quark distribution is smaller than up quark by a factor of
2 at some regions of $x$ and by the time $x$ reaches down to
$10^{-4}$,we have $x\bar{s}=x\bar{u}$. As for the charm quark content of
proton, the picture is less clear. Early treatments of heavy
parton distributions assumed that for $Q^{2}> m^{2}_{Q}$, the
heavy quark, $Q$, should be considered as massless. At the opposite extreme,
the heavy quark has never been regarded as part of the nucleon sea but
produced perturbatively through photon-gluon fusion. The number of flavors
remain fixed, regardless of $Q^{2}$. This treatment is used in GRV 94 parton
distribution\cite{7}. Both schemes have their own deficiency. For
example, in the former scheme, the heavy quark should not be
treated as massless for $Q^{2}\ge m^{2}_{Q}$ and in the latter
scheme one cannot incorporate large logarithms at $Q^{2}>>m^{2}_{Q}$.
CTEQ \cite{8} on the other hand, have used an interpolation, which produces
relative features of both schemes. In this treatment, the heavy quarks are
essentially produced by photon-gluon fusion when $Q^{2}\approx
m^{2}_{Q}$ and considered as massless quark when $Q^{2}>>
m^{2}_{Q}$. In our treatment, however, since we are dealing with
low $x$-values, we will assume $z\bar{s}=z\bar{u}$, and
$z\bar{c}=\alpha z\bar{s}$, where $\alpha$ is a factor taken to be
equal to the ratio of the strange and charm quark masses when $Q^{2}<
m^{2}_{charm}$; and $\alpha=1$ for $Q^{2}> m^{2}_{charm}$. Although
such an undertaking is not free of ambiguity, however, that does not
change our qualitative arguments regarding the normalization of the data.\\
Equations (1-3) completely determine the structure of a valon. The structure
function of any hadron can be written as the convolution of the structure
function of a valon with the valon distribution in the hadron:
\begin{equation}
F_{2}^{h}(x,Q^2)=\sum_{CQ}\int_{x}^{1}dy G_{\frac{CQ}{h}}(y)F^{CQ}_{2}(\frac{x}{y},Q^2)
\end{equation}
where summation runs over the number of CQ's in a particular hadron. $F_{2}^{CQ}(z, Q^{2})$
denotes the structure function of a CQ (U, D, $\bar{U}$, $\bar{D}$, etc.), as
given in equation (1), and $G_{\frac{CQ}{h}}(y)$ is the probability
of finding a valon carrying momentum fraction $y$ of the hadron. It is independent
of the nature of probe and its $Q^{2}$ value. Following \cite{4}, \cite{6},
and \cite{9}, for the case of pion we have:
\begin{equation}
G_{CQ/pion}(y)=\frac{1}{B(\mu +1, \nu +1)}y^{\mu}(1-y)^{\nu}
\end{equation}
where $G_{\frac{CQ}{pion}}$ is the U-valon distribution of in $\pi^{+}$ as well
as the D-valon distribution in $\pi^{-}$. Similar expression for
$G_{\bar{CQ}/pion}$, anti-valon distribution in a pion, is obtained by
interchanging $\mu\leftrightarrow \nu$. In the above equation $B(i,j)$ is
Euler $\beta$ function and $G_{\frac{CQ}{h}}(y)$ are non-invariant
distributions satisfying the following number and momentum sum
rules:
\begin{equation}
\int^{1}_{0}G_{\frac{CQ}{h}}(y) dy=1 \hspace{2cm}\sum_{CQ}\int^{1}_{0}yG_{\frac{CQ}{h}}(y) dy=1
\end{equation}
For pion, the numerical values are: $\mu=0.01$, $\nu =0.06$. We
have, however, tried a range of values for $\mu$ and $\nu$ and did
not find much sensitivity on $F_{2}^{\pi}$ data against this
variations. In \cite{4} \cite{10} it is estimated that
$\mu=\nu=0.$, which is very close to the values that are used in
this paper. The flatness or almost flatness of the valon
distribution in pion is attributed to the fact that the valons are
more massive than the pion, so they are tightly bound. From
$SU(2)$ symmetry one should expect that $\mu=\nu$. In our
calculation, $\mu$ is slightly different from $\nu$, indicating a
small violation of $SU(2)$ symmetry. This violation is very small
and the data on $F_{2}^{\pi}$ is not sensitive enough to make a
large difference. Significant asymmetry is observed in proton sea
and its implications are discussed in Ref. [6] in the context of the valon
model.   \\
The pion has two valons (or constituent quarks), for example,
$\pi^{+}$ has a U and a $\bar{D}$, therefore, the sum in equation
(4) has only two terms. Parton distributions in a pion, say
$\pi^{+}$, is obtained as:
\begin{equation}
u_{val.}^{\pi^{+}}(x,Q^2)=\int^{1}_{x}\frac{dy}{y}G_{\frac{U}{\pi^{+}}}(y)
u_{\frac{val.}{U}}(\frac{x}{y}, Q^{2})
\end{equation}
\begin{equation}
\bar{d}_{val.}^{\pi^{+}}(x, Q^{2})=\int^{1}_{x}G_{\frac{\bar{D}}{\pi^{+}}}(y)
\bar{d}_{\frac{val.}{\bar{D}}}(\frac{x}{y},Q^{2})\frac{dy}{y}
\end{equation}
\begin{equation}
q_{\frac{sea}{\pi^{+}}}(x,Q^{2})=\int^{1}_{x}\frac{dy}{y} G_{\frac{U}{\pi^{+}}}(y) q_{\frac{sea}{U}}(\frac{x}{y},Q^{2}) +
\int^{1}_{x}\frac{dy}{y} G_{\frac{\bar{D}}{\pi^{+}}}(y) q_{\frac{sea}{\bar{D}}}(\frac{x}{y},Q^{2})
\end{equation}
Similar relations can be written for $\pi^{-}$ and $\pi^{0}$. In
the above equations the subscripts $U$ and $\bar{D}$ are the two
valon types in $\pi^{+}$. Equations (7, 8, 9) along with equation
(4) completes the evaluation of the pion structure function. In
figures (1) we present $F_{2}^{\pi}(x, Q^2)$, as calculated above, for the
fixed $Q^{2}$'s corresponding to the ZEUS data of Ref.[1].
\section{DISCUSSION AND THE DATA}
In the previous section we outlined the procedure for calculating
$F_{2}^{\pi}(x, Q^2)$. The results are shown in figures (1) by the
square points. From the figures, it is evident that for smaller $x$ and
lower $Q^{2}$ values, the results of the model calculations are closer to
the additive quark model normalization of the ZEUS Coll. data (see
figure 19 of Ref.[1]). As we move towards the large $x$ values
the calculated structure function decreases and gets closer to the
effective flux normalization of the data. If we are to trust in 
the valon model results, which provides a very good description for the
wealth of proton structure function data as well as other hadronic
processes, we can conclude that the two normalizations used by ZEUS may
be relevant to different kinematical regions and lends support to the
assessment made by the ZEUS collaboration that the final results will
lie between the two options[3].
It is true that the valon model resembles the additive quark model in that
the contributions from each valon are added up. But here we are mainly
dealing with the parton content of each valon, which is derived from the
perturbative QCD. The valon distribution in pion serves only as a
phenomenological mimic of the pion wave function.
we further note that in the above calculation none of the
ZEUS data for pion structure function is used and no data fitting
is performed.  \\
ZEUS Coll. makes the observation that there is a simple
relationship between the proton structure function and the
effective pion flux normalization of the pion structure function
(see figure 18 of Ref. [1]); namely
\begin{equation}
F_{2}^{\pi(EF)}(x,Q^2)\approx kF_{2}^{p}(x,Q^2)
\end{equation}
with the proportionality constant, $k=0.361$. We have calculate the right-hand
side of Eq. (10) in the valon model and compared it with the effective flux
normalization, $F_{2}^{\pi(EF)}(x,Q^{2})$, in the left hand side. The copmarison
is presented in figures (1). As one can see from the figures, the relationship
holds rather well at all $Q^{2}$ values. To avoid any misleadings, we emphasizs
that the direct calculation of $F_{2}^{\pi}(x,Q^{2})$ in the valon model (Square
points in figures (1)) is
different from $F_{2}^{\pi(EF)}(x,Q^{2})$ and hence, does not support
the effective flux normalization of $F_{2}^{\pi}$. In other words, our
finding merely states that if we scale $F_{2}^{p}$ by a factor of $k=0.37$
we arrive at equation (10). Figures (1) also indicates that the above
relationship holds rather well at lower $Q^{2}$ values. As we move to higher
$Q^{2}$ this relationship at the lowest $x$-value gets blurred, but at
large $x$ and high $Q^{2}$ it continues to hold.\\
Since our model produces very good fit to the proton structure function data
in a wide range of both $x=[10^{-5}, 1]$ and $Q^{2}=[0.45, 10000]$ $Gev^{2}$,
we have also attempted to investigate the relation:
\begin{equation}
F_{2}^{\pi}(x, Q^{2})\approx \frac{2}{3}F_{2}^{p}(\frac{2}{3}x, Q^{2})
\end{equation}
which is based on color-dipole BFKL-Regee expansion and
corresponds to the ZEUS's additive quark model normalization. We have
calculated both sides of the equation (11) in the valon model. The results 
are shown by the solid lines and square points in Figures (1). Although We
get similar results as in Fig.19 of Ref.[1], but this does not say much about the
pion structure function data, because the additive quark model normalization
of the data is based on the above equation. It only restates that our model,
indeed, correctly produces proton structure function. The main
ingredient of our model is the partonic content of the constituent quark
which is calculated based on QCD dynamics. Convolution of this structure
with the constituent quark distribution in the pion appears, to some extent,
to give support for the additive quark model normalization of
the pion structure function data. In fact, we agree with Ref. [3]
that the final results should be somewhere between the two normalizations
used by ZEUS, being much closer to the additive quark model scheme than the
pion flux scheme. \\
It is worth to note that the following relationship also holds very
well between ZEUS's pion flux normalization data and proton
structure function:
\begin{equation}
F_{2}^{\pi(EF)}(x, Q^{2})= \frac{1}{3}F_{2}^{p}(\frac{2}{3}x, Q^{2})
\end{equation}
This relationship is essentially the same as Eq. (10), except for the factor
$\frac{2}{3}$ in front of $x$ that makes a small correction to the factor
$k$ of Eq. (10). In Figure (2) we have presented the right-hand sides of both
equations (10) and (12) along with the effective flux normalization data from
Ref.[1] at a typical value of $Q^{2}=15 GeV^{2}$. The same feature is also
prevalent for the other values of $Q^{2}$. \\
Both Eqs. (11) and (12) indicate that, regardless of the choice of normalization
scheme, the valence structure of proton, as compared to pion, is shifted to
the lower $x$ by a factor of $\frac{2}{3}$, So that valence $x$ in proton
corresponds to $\frac{2}{3}x$ in pion. \\
ZEUS has observed that the rate of neutron production in
photo-production process, in comparison to that of $pp$ collision,
drops to {\it{half}} and from this observation ZEUS Coll. has
concluded that $\sigma_{\gamma \pi} \approx
\frac{1}{3}\sigma_{\gamma p}$ whereas one expects to get a rather
$\frac{2}{3}$, both from Regge factorization and the additive
quark model[2]. This poses a problem on the understanding of the
dynamics of the interaction. A tentative resolution is that the
discrepancy can be resolved if we suppose that in the process each
valon interacts independently. That is the impulse approximation.
Suppose that $\delta$ denotes the number of valons in the target
proton that suffers a collision and $\delta_{i}$ denotes the
number of collisions that $i^{th}$ valon of the projectile
encounters. If we define the integer $\sigma=\sum_{i}\delta_{i}$,
then we will have $\delta \leq \sigma\leq 3 \delta$. Furthermore,
let $P_{\delta}(\sigma)$ represents the probability that out of
$\delta$ independent collisions that target valons encounter,
$\sigma$ valonic collisions occur. If $p$ is the probability that
either of the other valons in the projectile also interact, then
the probability for $\delta$ collisions will have a binomial
distribution having $\sigma - \delta$ valonic collision by $i=2$
and $3$ valons out of a maximum $2\delta$ possible such
collisions\cite{11}. Now, for a real photon, we can assume that it
may fluctuate into mesons with two valons. if we denote the number
of possible valonic collisions in $\gamma-p$ interaction by
$\delta^{'}$ then the mean number of such  collisions will be
$2\delta^{'}p$ whereas in $pp$ collisions it will be $3\delta p$.
The observed reduction in the rate of neutron production in two
processes and the conclusion that
$\sigma_{\gamma\pi}\approx\frac{1}{3}\sigma_{pp}$ implies that
$\frac{\delta^{'}}{\delta}=\frac{1}{2}$. That is, there are twice
as many collisions in $pp$ interactions than in photo-production.
\section{conclusion}
We have demonstrated that the assumption of the existence of a
basic structure in hadrons is a reasonable model to investigate
the hadronic structure. Pion structure function measurement by
ZEUS Coll. provides additional tests and further validation of the
Model. We have presented a resolution to the issue of the
normalization of the data. Our results are based on QCD dynamics and
suggest that the correct normalization of $F_{2}^{\pi}$ is closer
to the additive quark model normalization for low $x$ region. The
observed reduction in the rate of neutron production in
photo-production as compared to $pp$ collision is accounted for
and concluded that there are twice less valonic collisions in
photo-production than in $pp$ collision. Furthermore, it appears
that there are simple relations among the structure functions of
hadrons; namely they are proportional and the proportionality
ratio seems to depend on the normalization scheme chosen.

\section{Acknowledgment}
I am grateful to Prof. Garry Levman for useful discussions and providing me
with the experimental data. I am also grateful to Professor Rudolph C. Hwa, who
kindly has read the manuscript and made valuable comments, and pointed out an
error in Eq. (4).

\end{document}